\documentclass[aps,a4paper, preprint, superscriptaddress,preprintnumbers,floatfix,nofootinbib,amsmath,amssymb]{revtex4-1}

\usepackage{url}
\usepackage{hyperref}
\usepackage{color}
\usepackage{cancel}
\usepackage{cleveref}
\usepackage{soul}
\usepackage[normalem]{ulem}

\usepackage{amstext,amssymb}
\usepackage{amsmath}
\usepackage{graphicx}
\usepackage{diagbox}
\usepackage{slashbox}
\usepackage{slashbox}
\usepackage{booktabs}
\usepackage{xspace}
\usepackage{color}
\usepackage{units}
\usepackage[T1]{fontenc}
\usepackage{amsmath,bm}
\usepackage{nicefrac}
\usepackage{slashed} 
\usepackage{multirow}
\newcommand{\be}{\begin{equation}}
\newcommand{\ee}{\end{equation}}
\newcommand{\bea}{\begin{eqnarray}}
\newcommand{\eea}{\end{eqnarray}}

\usepackage{slashed}

\newcommand{\nua}[1]{\ensuremath{\rlap{\kern-2.5pt\ensuremath{\overset{\scriptscriptstyle(-)}{\phantom{\nu}}}}{\ensuremath{{\nu}_{#1}}}}\xspace}

\definecolor{brickred}{rgb}{0.8, 0.25, 0.33}
\definecolor{brightcerulean}{rgb}{0.11, 0.67, 0.84}
\definecolor{brown(traditional)}{rgb}{0.59, 0.29, 0.0}
\begin{document}
\title{Sensitivities of Long and Medium Baseline Experiments to Sterile Neutrinos and Non-Standard Interactions}

\author{Sambit Kumar Pusty}
\email{pustysambit@gmail.com}
\affiliation{School of Physics,  University of Hyderabad, Hyderabad - 500046,  India}

\author{Trisha Guin}
\email{guin.trisha@gmail.com}
\affiliation{School of Physics,  University of Hyderabad, Hyderabad - 500046,  India}

\author{Rukmani Mohanta}
\email{rmsp@uohyd.ac.in}
\affiliation{School of Physics,  University of Hyderabad, Hyderabad - 500046,  India}

\begin{abstract}
    In this work, we present for the first time,  the simultaneous analysis of the non-standard neutrino interactions (NSI) and a light sterile neutrino in the 3+1 framework for the future long- and medium-baseline experiments DUNE and MOMENT, respectively. We show that treating the two sectors separately yields over optimistic sensitivities.   The simultaneous presence of sterile-neutrino mixing and complex NSI  weakens the  $|\varepsilon_{e\mu}|$ sensitivity by a factor of $\sim (2-3)$ with the dominant degradation associated with the additional parameter freedom in the flavor-changing  NSI sector. DUNE alone fails to achieve $5\sigma$ CP-violation discovery, once NSI phases are marginalized over. The DUNE+MOMENT combination recovers $>5\sigma$ sensitivity over a wide range of $\delta_{\rm CP}$ space, constrains $\sin^2\theta_{24}\lesssim10^{-2}$ and $|\varepsilon_{e\mu}|\lesssim0.025$ at $95\%$~C.L., and resolves degeneracies that are intractable for either experiment alone. Our results establish that a multi-baseline strategy combining matter-rich and near-vacuum baselines is necessary for robust parameter extraction in the  simultaneous presence of two new physics scenarios. It is also important to perform simultaneous new physics analyses to reliably interpret future precision neutrino oscillation data.
\end{abstract}

\maketitle
\flushbottom

\section{Introduction}
Neutrinos are one of the most fascinating particles in the universe. They have tiny masses,  barely interact with matter, transmute from one flavor to another during their propagation through space and this phenomenon of flavor conversion is known as neutrino oscillation \cite{Giunti:2007ry,Bilenky:1998dt}. The discovery of neutrino oscillations from the atmospheric (Super-Kamiokande) \cite{KAJITA201614} and the Solar (SNO) \cite{GRAHAM2003C556} neutrino experiments led to the Physics Nobel Prize in 2015. The three flavor eigenstates of neutrinos ($\nu_e,\nu_\mu,\nu_\tau$) are related to the corresponding    mass eigenstates ($\nu_1,\nu_2,\nu_3$) through the unitary Pontecorvo-Maki–Nakagawa–Sakata (PMNS) mixing matrix,  
parametrized in terms of three mixing angles ($\theta_{12},\theta_{13},\theta_{23}$) and one CP-violating Dirac phase ($\delta_{CP}$) \cite{Giganti:2017fhf}. 
The major unknowns in neutrino oscillation at present are: the mass hierarchy, i.e., whether $\Delta m_{31}^2>0$ or $\Delta m_{31}^2<0$, usually referred to as normal/inverted hierarchy, the value of the CP-violating phase $\delta_{CP}$, and the octant of the atmospheric mixing angle $\theta_{23}$. 
In this modern precision era of neutrino physics, the results from the two leading long-baseline experiments on the  CP-violating phase $\delta_{CP}$ appear to be inconsistent. The T2K experiment, using a 295 km baseline with its $L/E$ optimized for the atmospheric oscillation maximum, reports the best-fit value as $\delta_{CP}\sim1.4\pi$, disfavoring the CP-conserving values at $3\sigma$ ~\cite{T2K:2019bcf}. The NOvA experiment with 810 km baseline, on the other hand, finds $\delta_{CP}\sim0.8\pi$ ~\cite{NOvA:2021nfi}, which appears to be in tension with the T2K result, when both the experiments analyzed their data within the standard three-flavor framework. While systematic uncertainties between the two experiments can partly account for this gap, the possibility that the discrepancy could instead be a signal of beyond-Standard-Model (BSM) physics, either in the form of Non-Standard Interactions (NSI) or possibly due to new additional states, i.e., sterile neutrinos.  Recent studies~\cite{Chatterjee:2020kkm,Chatterjee:2024kbn} have shown that NSI with $|\varepsilon_{e\mu}| \sim |\varepsilon_{e\tau}| \sim 0.1$ can alleviate this discrepancy, with subsequent analyses supporting this possibility at the $\sim 1.8\sigma$ level. An alternative pseudo-sterile interpretation involving novel matter potentials has also been proposed in Ref~\cite{Chatterjee:2026ctb}. Nevertheless, a comprehensive study of how NSI and sterile neutrinos simultaneously affect future experiment sensitivities remains limited.
As neutrino oscillation experiments are advancing into the era of precision measurements, it would be interesting to quantitatively investigate the effects of these two scenarios simultaneously for correctly interpreting future precision oscillation data.\\

The possibility of light sterile neutrinos has been extensively discussed in light of several experimental anomalies. The LSND~\cite{LSND:2001aii} and MiniBooNE~\cite{MiniBooNE:2020pnu} experiments observed an excess of electron (anti)neutrino events in predominantly muon (anti)neutrino beams, while the reactor antineutrino anomaly~\cite{Mention:2011rk,Giunti:2021kab} and Gallium anomaly~\cite{Acero:2007su,GALLEX:1997lja} reported deficits in the observed electron (anti)neutrino fluxes compared to theoretical predictions. The minimally well-motivated explanation for the combined LSND, MiniBooNE, reactor, and Gallium anomalies is the existence of a fourth sterile neutrino state $\nu_s$ with a mass-squared splitting $\Delta m^2_{41}\sim1~{\rm eV}^2$ \cite{Dasgupta:2021ies}. It couples to the active sector through small but nonzero mixing angles.
 In the minimal extension of the Standard Model, we have only one new sterile state, generally referred to  as 3+1 active-sterile framework with three additional mixing angles: ($\theta_{14},\theta_{24},\theta_{34}$), one mass squared difference ($\Delta m^2_{41}$) and two new phases: ($\delta_{24},\delta_{34}$). Although the recent MicroBooNE result~\cite{MicroBooNE:2025nll} has excluded a large fraction of the light sterile neutrino parameter space favored by LSND and MiniBooNE, they do not rule out the entire 3+1 allowed region, and the sterile neutrino scenario still remains well-motivated both experimentally and theoretically ~\cite{Agarwalla:2018nlx,Choubey:2017cba,Choubey:2017ppj,Haba:2018klh,Coloma:2017ptb,Berryman:2015nua, Majhi:2019hdj,Sharma:2023jzg,Cabrera:2025qcs,Parveen:2024bcc,Acero:2022wqg,universe12040105}. Still persistent several unresolved anomalies, the possibility of more complex sterile-neutrino scenarios, connections to neutrino mass generation, dark matter, as well as cosmology continue to drive this area an active field of research.

In parallel, Non-Standard Interactions (NSI) of neutrinos with matter, arising from effective four-fermion operators beyond the standard model, provide an independent and well-motivated BSM framework ~\cite{Wolfenstein:1977ue,Guzzo:1991hi,Valle:1987gv,Miranda:2015dra,Ohlsson:2012kf,Farzan:2017xzy,Proceedings:2019qno,Biggio:2009nt,Esteban:2018ppq,Coloma:2019mbs,Coloma:2023ixt,Dutta:2020che,Super-Kamiokande:2011dam,IceCubeCollaboration:2021euf,Krishnamoorthi:2025efw,KM3NeT:2024pte,MINOS:2013hmj,NOvA:2024lti,Borexino:2019mhy,BOREXINO:2026owb,Singha:2021jkn,Ghosh:2017lim,ESSnuSB:2025vsf,Ohlsson:2012kf,Liao:2016orc}. These NSI are generally assumed to be generated from the interactions of standard model fermions with new heavy  mediators and are of great phenomenological interest as their presence  can alter the dynamics  of the neutrino flavor conversion in matter.

In this work, we intend to  study the combined effect of NSI and the 3+1 active-sterile mixing scenarios, for the first time to the best of our knowledge. In addition to the standard matter effect, we consider that the propagation of neutrinos in matter is influenced
by NSI via neutral-current (NC) interactions, while  the charged-current (CC) NSI affecting the production and detection processes of the neutrinos at the source and detector are neglected. 
The effective Lagrangian for neutrino NSI incorporating matter effect is presented in Sec.~\ref{NSI_theory}.
During the propagation, NSI modifies the effective matter Hamiltonian, shifting both the amplitude and energy dependence of oscillation probabilities in such a way  that it can mimic the effects on mixing angles, mass orderings, and CP phases~\cite{Miranda:2015dra,Farzan:2017xzy}. In particular, the NSI rescales the effective matter potential seen by neutrinos, making it exceptionally difficult to isolate from standard oscillation parameters using a single experiment with specific baseline.

One of the critical and largely unexplored domains in neutrino sector is the simultaneous presence of both NSI and sterile-neutrino effects. Most of the existing analyses treat these two BSM scenarios independently: studies of sterile neutrinos typically assume standard interactions with matter~\cite{Berryman:2015nua,Blennow:2016jkn}, while NSI analyses in the same context assume non-standard interactions of three active flavors without any sterile admixture. This decoupled approach is insufficient because the two frameworks introduce overlapping parameter degeneracies. For example, the effective matter potential with NSI, i.e.,  $V^{NSI}=\sqrt 2G_F N_e(1+\varepsilon_{ee})$ and the sterile CP phase $\delta_{24}$ can jointly mimic the effect of $\delta_{CP}$ in the appearance channel. A simultaneous treatment is therefore necessary to correctly interpret data from next-generation experiments. Thus, we address this gap by performing a comprehensive simultaneous analysis of neutral-current NSI and the 3+1 active-sterile mixing framework for the two next-generation long-baseline experiments: DUNE and MOMENT, for the first time.
Both  DUNE and MOMENT experiments have unique features to explore beyond the standard model sub-leading effects of sterile neutrinos and non-standard interactions. The detailed technical configurations of these two experiments are given in Sec.~\ref{Experimental_Setup}. These two upcoming experiments with different detectors, different baselines, and different energy ranges are expected to provide deeper insights into new physics exploration.

The outline of the paper is as follows. In Sec.~\ref{Theoretical}, we discuss the theoretical framework for sterile neutrinos and NSI effects on neutrino oscillations. Sec.~\ref{Experimental_Setup} presents the details of the experimental setups for the DUNE and MOMENT experiments, as well as the simulation details used in this study. The results are described in Sec.~\ref{results}, and finally, in Sec.~\ref{coclusion} we summarize the conclusions of this study.
\section{Theoretical Framework}
\label{Theoretical}
\subsection{Standard Three-Flavor Oscillations}

Neutrino flavor eigenstates $|\nu_\alpha \rangle$  are related to the mass eigenstates $|\nu_i\rangle$  through the  unitary PMNS mixing matrix $U$ as,
\begin{equation}
|\nu_\alpha \rangle=\sum_{i=1}^{3}
U_{\alpha i}|\nu_i \rangle,
\qquad
\alpha=e,\mu,\tau.
\end{equation}
The propagation of neutrinos in vacuum is described by the Schrödinger-like evolution equation
\begin{equation}
i\frac{d}{dt}|\nu_\alpha\rangle=H_{\rm vac}|\nu_\alpha \rangle,
\end{equation}
and the vacuum Hamiltonian is given by
\begin{equation}
    H_{\rm vac} = \frac{1}{2E}\,U_{\rm PMNS}\,M^2\,U_{\rm PMNS}^\dagger,\quad{\rm with} \quad
    \quad
    M^2 = \mathrm{diag}(0,\,\Delta m_{21}^2,\,\Delta m_{31}^2),
\end{equation}
where $\Delta m_{ij}^2 \equiv m_i^2 - m_j^2$ and $E$ is the neutrino energy.
The probability for a neutrino produced in flavor state $\alpha$ to be detected in flavor state $\beta$ after propagating a distance $L$ is
\begin{equation}
P_{\alpha\beta}=\left|\langle \nu_\beta|
e^{-iH_{\rm eff}^{3 \nu}L} | \nu_\alpha \rangle
\right|^2.
\end{equation}
When neutrinos propagate through matter, coherent forward scattering with electrons modifies the oscillation pattern. This phenomenon is known as the Mikheyev-Smirnov-Wolfenstein (MSW) effect. So the effective Hamiltonian becomes
\begin{equation}
H_{\rm eff}^{3 \nu}=H_{\rm vac}
+
V_{\rm mat},
\end{equation}
where $V_{\rm mat}=\sqrt{2}G_F N_e~ \rm diag(1,0,0)$. 
Here, $G_F$ is the Fermi constant and $N_e$ represents the electron number density of the medium.
\subsection{Non-Standard Interactions (NSI)}
\label{NSI_theory}
There are many extensions of the Standard Model in the literature that predict new additional neutrino interactions with matter. These NSI effects can be described through effective four-fermion operators \cite{Miranda:2015dra},
\begin{equation}
\mathcal{L}_{\rm NSI}=
-2\sqrt{2}G_F
\varepsilon_{\alpha\beta}^{ff'C}
\left(
\bar{\nu}_{\alpha L}\gamma^\mu \nu_{\beta L}
\right)
\left(
\bar{f_C}\gamma_\mu  f'_C
\right).
\end{equation}
Here,  $\varepsilon_{\alpha\beta}^{ff'C}$ is the NSI parameter that defines the strength of non-standard interaction between the $\alpha$ and $\beta$ flavored leptons, $f,f'=e,u,d$, and $C=L,R$ (chiral projection operators). For $f=f'$, the interaction is referred to as neutral-current like NSI that affects the neutrino propagation in matter. \cite{Ohlsson:2012kf,Farzan:2017xzy}  For $\varepsilon_{\alpha\beta} \to 0$, standard interactions can be restored.

For neutrino propagation in matter, these interactions modify the matter potential.  The NSI potential is Hermitian, i.e., $\varepsilon_{\alpha\beta} = \varepsilon_{\beta\alpha}^*$;  the matrix therefore has 3 real diagonal and 3 complex off-diagonal parameters. The full matter potential in the three-flavor case becomes
\begin{equation}
V_{\rm NSI}=
\sqrt{2}G_F N_e
\begin{pmatrix}
\varepsilon_{ee} & \varepsilon_{e\mu} & \varepsilon_{e\tau}\\
\varepsilon_{\mu e} & \varepsilon_{\mu\mu} & \varepsilon_{\mu\tau}\\
\varepsilon_{\tau e} & \varepsilon_{\tau\mu} & \varepsilon_{\tau\tau}
\end{pmatrix}.
\end{equation}
Thus, the  effective Hamiltonian, including both standard and non-standard matter effects, is given by                             
\begin{equation}
H^{3\nu}_{\rm NSI}=
\frac{1}{2E}
U_{\rm PMNS}
M^2
U_{\rm PMNS}^{\dagger}
+
V_{\rm mat}
+
V_{\rm NSI}
=
\frac{1}{2E}
U_{\rm PMNS}
M^2
U_{\rm PMNS}^{\dagger}
+
\sqrt{2}G_F N_e
\left(
I+\varepsilon
\right).\label{eq:8}
\end{equation}
The identity term on the right-hand side of (\ref{eq:8}) represents the standard MSW potential, and the $\varepsilon$ matrix depicts the NSI effect. The diagonal NSI parameters $\varepsilon_{\alpha\alpha}$ modify the effective matter potential for each flavor, while the off-diagonal $\varepsilon_{\alpha\beta}$  ($\alpha\neq\beta$) induce flavor-changing interactions during propagation. 

\subsection{Sterile Neutrino Framework}

Motivated by several experimental anomalies, the standard three-flavor neutrino framework can be extended by adding one new sterile neutrino state, $\nu_s$. The resulting 3+1 framework contains four flavor eigenstates  $(\nu_e, \nu_\mu, \nu_\tau, \nu_s)$ and four mass eigenstates  $(\nu_1, \nu_2, \nu_3, \nu_4)$, with $m_4 \gg m_{1,2,3}$. The vacuum Hamiltonian in the 3+1 framework is given as,
\begin{equation}
H_{\rm vac}^{3+1}=
\frac{1}{2E}
U_{4\times4}
\begin{pmatrix}
0 & 0 & 0 & 0 \\
0 & \Delta m_{21}^{2} & 0 & 0 \\
0 & 0 & \Delta m_{31}^{2} & 0 \\
0 & 0 & 0 & \Delta m_{41}^{2}
\end{pmatrix}
U_{4\times4}^{\dagger}\;.
\end{equation}
The extended mixing matrix $U_{4\times4}$ introduces three additional mixing angles,
$\theta_{14}, \theta_{24}, \theta_{34},$ and two additional CP-violating phases,
$\delta_{24}, \delta_{34}.$ 
Now, the $4\times4$ mixing matrix is parameterized as
\begin{equation}
U_{4\times4}=
\widetilde{R}_{34}
\widetilde{R}_{24}
R_{14}
R_{23}
\widetilde{R}_{13}
R_{12}\;,
\end{equation}
where $\widetilde{R}_{ij}$ ($R_{ij}$) denotes a complex (real) rotation matrix in the $(ij)$ plane containing the corresponding CP phase (with no CP phase).

Since sterile neutrinos do not participate in Standard Model weak interactions, the standard $4\times4$ matter potential is therefore

\begin{equation}
\label{eq:VSM_31}
V^{3+1}=
\begin{pmatrix}
V_{\rm CC}+V_{\rm NC} & 0 & 0 & 0 \\
0 & V_{\rm NC} & 0 & 0 \\
0 & 0 & V_{\rm NC} & 0 \\
0 & 0 & 0 & 0
\end{pmatrix},
\end{equation}
where
\begin{equation}
V_{\rm CC}=\sqrt{2}G_F N_e,
\qquad
V_{\rm NC}=-\frac{1}{\sqrt 2}G_F N_n.
\end{equation}
Unlike the three-flavor case, one cannot simply drop $V_{\rm NC}$ here, 
because it does not act uniformly on all four states: the sterile neutrino 
receives no contribution, while all active flavors do.
Hence, we can subtract the common factor $V_{\rm NC}\times\mathbf{I}_{4\times4}$ 
from Eq.~\eqref{eq:VSM_31} to obtain the physically equivalent effective form:
\begin{equation}
\label{eq:VSM_eff}
    V^{3+1} = 
    \mathrm{diag}\!\left(
        V_{\rm CC},\;
        0,\;
        0,\;
        -V_{\rm NC}
    \right)
    =
    V_{\rm CC}\,\mathrm{diag}(1,\;0,\;0,\;r),
\end{equation}
where
\begin{equation}
\label{eq:r_def}
    r \equiv -\frac{V_{\rm NC}}{V_{\rm CC}} \approx 0.5.
\end{equation}
For the Earth's mantle, the PREM  model~\cite{Dziewonski:1981xy} gives an average electron fraction  $Y_e \equiv N_e/(N_p+N_n)\approx0.494$, where $N_p$ and $N_n$ are the
proton and neutron number densities, corresponding to  $r = (1-Y_e)/(2Y_e)=0.512 \approx 0.5$.

\subsection{Combined Sterile Neutrino and NSI Framework}

The most general scenario considered in this work incorporates both sterile neutrino mixing and NSI simultaneously. In the 3+1 framework, the NSI potential extended to the sterile sector takes the form
\begin{equation}
V_{\rm NSI}^{3+1}=
\sqrt{2}G_F N_e
\begin{pmatrix}
\varepsilon_{ee} &
\varepsilon_{e\mu} &
\varepsilon_{e\tau} &
\varepsilon_{es}
\\
\varepsilon_{\mu e} &
\varepsilon_{\mu\mu} &
\varepsilon_{\mu\tau} &
\varepsilon_{\mu s}
\\
\varepsilon_{\tau e} &
\varepsilon_{\tau\mu} &
\varepsilon_{\tau\tau} &
\varepsilon_{\tau s}
\\
\varepsilon_{s e} &
\varepsilon_{s\mu} &
\varepsilon_{s\tau} &
\varepsilon_{ss}
\end{pmatrix}.
\end{equation}
Thus, the complete Hamiltonian governing neutrino propagation can be written as follows,
\begin{equation}
\label{Hfull}
H=
\frac{1}{2E}
U_{4\times4}
M^2
U_{4\times4}^{\dagger}
+
V^{3+1}
+
V_{\rm NSI}^{3+1}\;,
\end{equation}
where $M^2 = \mathrm{diag}(0, \Delta m_{21}^2, \Delta m_{31}^2, \Delta m_{41}^2)$, and the total potential can be expressed as, 
\begin{equation}
\label{V_final}
    V=V^{3+1}+V_{\rm NSI}^{3+1}=V_{CC}
    \begin{pmatrix}
        1+\varepsilon_{ee} &
\varepsilon_{e\mu} &
\varepsilon_{e\tau} &
\varepsilon_{es}
\\
\varepsilon_{\mu e} &
\varepsilon_{\mu\mu} &
\varepsilon_{\mu\tau} &
\varepsilon_{\mu s}
\\
\varepsilon_{\tau e} &
\varepsilon_{\tau\mu} &
\varepsilon_{\tau\tau} &
\varepsilon_{\tau s}
\\
\varepsilon_{s e} &
\varepsilon_{s\mu} &
\varepsilon_{s\tau} &
\varepsilon_{ss}+r
    \end{pmatrix}\;.
\end{equation}

Eq.~\eqref{V_final} is the central theme of this work: 
the $(1,1)$ entry $1+\varepsilon_{ee}$ shows how NSI rescales the effective 
electron-neutrino potential, and the $(4,4)$ entry $r$ encodes the 
sterile-active potential information. Since both sectors modify the flavor-transition amplitudes, significant parameter degeneracies may arise, particularly among the combinations $\theta_{14},~\theta_{24}$ and $\varepsilon_{e\mu}$, as well as between the phases $\delta_{24}$ and $\phi_{e\mu}$. Understanding and resolving these degeneracies constitutes one of the primary objectives of the present work.


\section{Experimental Setup and Simulation Details}
\label{Experimental_Setup}
In this section, we describe the key features of the experimental configurations for the long-baseline DUNE and the medium-baseline MOMENT experiments. 
For the DUNE simulations, we use the official configuration files provided in the Technical Design Report (TDR)~\cite{DUNE:2021cuw}. The DUNE far detector consists of four liquid argon time projection chamber (LArTPC) modules, each with a fiducial mass of 10 kt, and is designed to operate with a 1.2 MW proton beam. The total data-taking period is assumed to be 13 years, equally divided between neutrino and antineutrino running modes, corresponding to 6.5 years in each mode. This results in a total exposure of 624 kt-MW-years, assuming an annual beam intensity of $1.1 \times 10^{21}$ protons on target (POT). Throughout the simulations, the matter density along the 1300 km baseline is assumed to be uniform, with a constant value of 2.848~g/cm$^3$.
The MOMENT experiment will have an estimated run-time of 10 years, equally split between neutrino and antineutrino modes. It has a 15 MW proton beam at 1.5 GeV energy, which will be equivalent to $1.1 \times 10^{24}$ protons on target (POT) per year. The density of Earth matter is estimated to be around 2.7 ${\rm g/cm^3}$. In addition, it has the advantage of probing eight different oscillation channels, which helps reduce systematic uncertainties and low beam-induced backgrounds.

The numerical simulations in this work are performed using the GLoBES (General Long Baseline Experiment Simulator)  package~\cite{Huber:2004ka,Huber:2007ji}. GLoBES is a software package designed to simulate and analyze long-baseline and reactor neutrino oscillation experiments. For the evaluation of experimental sensitivity, the Poisson log-likelihood statistical formula is defined as
\begin{equation}
    \chi^2_{\mathrm{stat}}
=
2 \sum_{i=1}^{n}
\left[
N_i^{\mathrm{test}}
-
N_i^{\mathrm{true}}
-
N_i^{\mathrm{true}}
\log\!\left(
\frac{N_i^{\mathrm{test}}}
     {N_i^{\mathrm{true}}}
\right)
\right],
    \end{equation}
 where $N^{\rm test}_i$ and $N^{\rm true}_i$ represent the test and true event rates, respectively. The subscript $i$ indicates the energy bins. The impact of systematic uncertainties is incorporated into our simulation using the pull method, where the corresponding nuisance parameters $\zeta$ (representing signal normalization and background normalization errors) are marginalized simultaneously in the $\chi^2$ minimization \cite{Fogli:2002pt,Huber:2006wb}. $1\sigma$ uncertainties on the standard oscillation parameters, as shown in Table \ref{osc_par}, are also incorporated through the input-error vector. We have implemented two types of systematic uncertainties: signal error and background normalization error, as shown in Table \ref{systematics}.
\begin{table}[htbp]
\centering

\begin{tabular}{|c|c|c|}
\specialrule{1.5pt}{0pt}{0pt}
~\textbf{Systematic Uncertainties} ~&~ \textbf{DUNE}~ &~ \textbf{MOMENT}~ \\
\hline

Signal $\nu_{\mu}$ & 5\% & 5\% \\
\hline

Signal $\nu_{e}$ & 2\% & 2.5\% \\
\hline

Background Normalization & (5--20)\% & 5\% \\
\specialrule{1.5pt}{0pt}{0pt}
\end{tabular}

\caption{Systematic uncertainties used in this work.}
\label{systematics}

\end{table}
The standard neutrino oscillation parameters are taken from the NuFIT global analysis \cite{Esteban:2024eli}. The sterile neutrino and NSI parameters are adopted from recent global analyses available in the literature. The full parameter space of Eq.~\eqref{Hfull} is extremely large. To make the problem tractable while retaining the most phenomenologically relevant effects, we adopt the following simplifications: We consider three benchmark values of the sterile neutrino mass-squared splitting, $\Delta m^2_{41}=0.01,~0.1,~\text{and}~ 1~\mathrm{eV}^2$, while fixing the active-sterile mixing angles: $\sin^2\theta_{14}=\sin^2\theta_{24}=0.0076$ \cite{Singha:2022btw}. For the $1~\mathrm{eV}^2$ analysis, $\Delta m^2_{41}=1~\mathrm{eV}^2$ is fixed in both the true and test hypotheses, while $\theta_{34}=0$ and $\delta_{34}=0$ are fixed in both. The mixing angles $\theta_{14}$ and $\theta_{24}$ are allowed to vary in the test hypothesis. The sterile phase $\delta_{24}$ is marginalized over its full  range: $\delta_{24}\in (0-360)^\circ$. The NSI parameters $\varepsilon_{ee}$ and $|\varepsilon_{e\mu}|$ are marginalized, together with the NSI phase $\phi_{e\mu}$ over its full range. In this work, we restrict our analysis to one flavor-diagonal NSI parameter $\varepsilon_{ee}$ and one complex flavor-changing off-diagonal NSI parameter $\varepsilon_{e\mu}=|\varepsilon_{e\mu}|e^{i\phi_{e\mu}}$ as representative parameters while setting all others to zero. The former modifies the effective electron-matter potential, while the latter directly affects the $\nu_\mu\rightarrow\nu_e$ appearance channel and introduces an additional CP phase. We emphasize that this choice does not imply that $\varepsilon_{e\tau}$ or sterile-sector matter couplings such as $\varepsilon_{ss}$ are unimportant; their inclusion would require a substantially larger parameter space and is beyond the scope of the present study.
 A summary of all the parameters used in this study is provided in the Table \ref{osc_par}. 

 \begin{table}[htbp]
\centering
\begin{tabular}{|c|c|c|}
\specialrule{1.5pt}{0pt}{0pt}
\textbf{Parameter} & \textbf{True Values $\pm1\sigma$} & \textbf{$3\sigma$ Range} \\
\hline

$\sin^2\theta_{12}$ &
$0.3088^{+0.0067}_{-0.0066}$ &
$0.2893$--$0.3295$ \\
\hline

$\sin^2\theta_{13}$ &
$0.02248^{+0.00055}_{-0.00059}$ &
$0.02064$--$0.02418$ \\
\hline

$\sin^2\theta_{23}$ &
$0.470^{+0.017}_{-0.014}$ &
$0.435$--$0.584$ \\
\hline

$\delta_{\rm CP}\,[^\circ]$ &
$212^{+26}_{-36}$ &
$125$--$365$ \\
\hline

$\Delta m^2_{21}\,[10^{-5}\,\mathrm{eV}^2]$ &
$7.537^{+0.094}_{-0.10}$ &
$7.236$--$7.823$ \\
\hline

$\Delta m^2_{31}\,[10^{-3}\,\mathrm{eV}^2]$ &
$+2.51^{+0.021}_{-0.020}$ &
$2.450$--$2.576 $ \\
\specialrule{1.5pt}{0pt}{0pt}

$\sin^2\theta_{14}$ & $0.0076$ & -- \\
\hline

$\sin^2\theta_{24}$ & $0.0076$ & -- \\
\hline

$\sin^2\theta_{34}$ & $0$ & -- \\
\hline

$\delta_{24}\,[^\circ]$ & $0$ & -- \\
\hline

$\delta_{34}\,[^\circ]$ & $0$ & -- \\
\hline

$\Delta m^2_{41}\,[\mathrm{eV}^2]$ &
$0.01,\;0.1,\;1$ &
-- \\
\specialrule{1.5pt}{0pt}{0pt}
$\varepsilon_{ee}$ & $0.2$ & --\\
\hline
$\varepsilon_{e\mu}$ & $0.02$ & --\\
\hline
$\phi_{e\mu}\,[^\circ]$ & $90$ & --\\

\specialrule{1.5pt}{0pt}{0pt}
\end{tabular}
\caption{The true oscillation parameters are taken from the NuFit 6.1~\cite{Esteban:2024eli} global fit and are listed in the first section. The sterile neutrino parameters are given in the second section, while the NSI parameters are presented in the third section. All remaining parameters are fixed to their standard values.}
\label{osc_par}
\end{table}

\section{Results}
\label{results}
Before presenting the $\chi^2$ sensitivity analyses, it is crucial to first examine how sterile neutrino mixing and NSI modify the underlying oscillation probabilities. Since the expected event rates in long-baseline experiments depend directly on the neutrino oscillation probabilities, understanding these probability-level effects provides valuable physical insight into the origins of the sensitivities and parameter degeneracies observed in subsequent statistical analyses. In particular, sterile neutrinos introduce additional oscillation frequencies through the new mass-squared splitting $\Delta m_{41}^2$, while propagation NSI modify the effective matter potential experienced by neutrinos during propagation. When both effects are present simultaneously, their interplay can lead to constructive or destructive interference, significantly altering the oscillation pattern depending on the neutrino energy and experimental baseline.

We then present the sensitivities obtained from the $\chi^2$ analyses on the sterile-neutrino and NSI parameter spaces in Sec.~\ref{sterilebound} and \ref{nsibound}. To investigate the interplay between these two classes of new physics, we perform a dedicated analysis in which each sector is studied both independently and in the presence of the other. For the sterile-neutrino scenario, we derive contour plots in the $\sin^2\theta_{14}-\sin^2\theta_{24}$ plane for three representative benchmark values of $\Delta m_{41}^2$. The resulting constraints are obtained under two distinct assumptions: (i) a pure $3+1$ sterile framework without NSI effects and (ii) a sterile$+\mathrm{NSI}$ framework in which the NSI parameters are included.
Similarly, we investigate the constraints on the NSI parameter space by considering one diagonal NSI parameter ($\varepsilon_{ee}$), and one off-diagonal parameter ($\varepsilon_{e\mu}$). The corresponding exclusion regions are obtained in two complementary scenarios. In the first scenario, the bounds are derived assuming the standard three-flavor framework with only NSI effects, whereas in the second scenario, the sterile-neutrino parameters are incorporated simultaneously. The comparison between these two cases provides a direct measure of the impact of active-sterile mixing on the determination of NSI parameters. Taken together, these analyses enable a systematic study of the correlations and degeneracies between sterile-neutrino mixing and non-standard interactions and reveal how the presence of one kind of new physics can affect the inferred constraints on the other.

\subsection{Probability-Level Analysis}

\begin{figure}[h]
\begin{center}
\includegraphics[width=78mm, height=54mm]{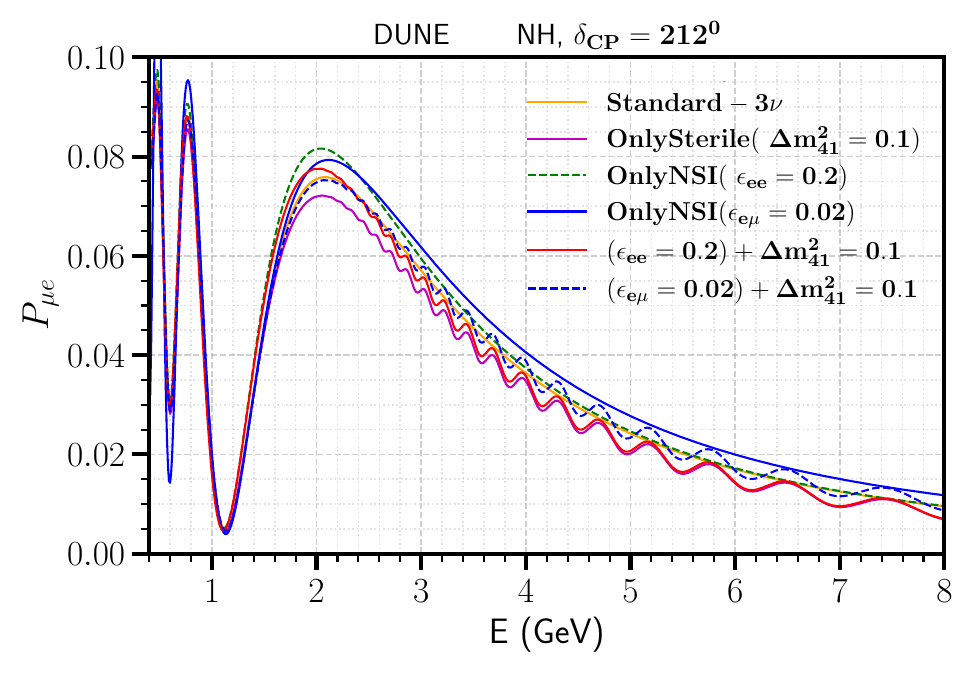}
\includegraphics[width=78mm, height=54mm]{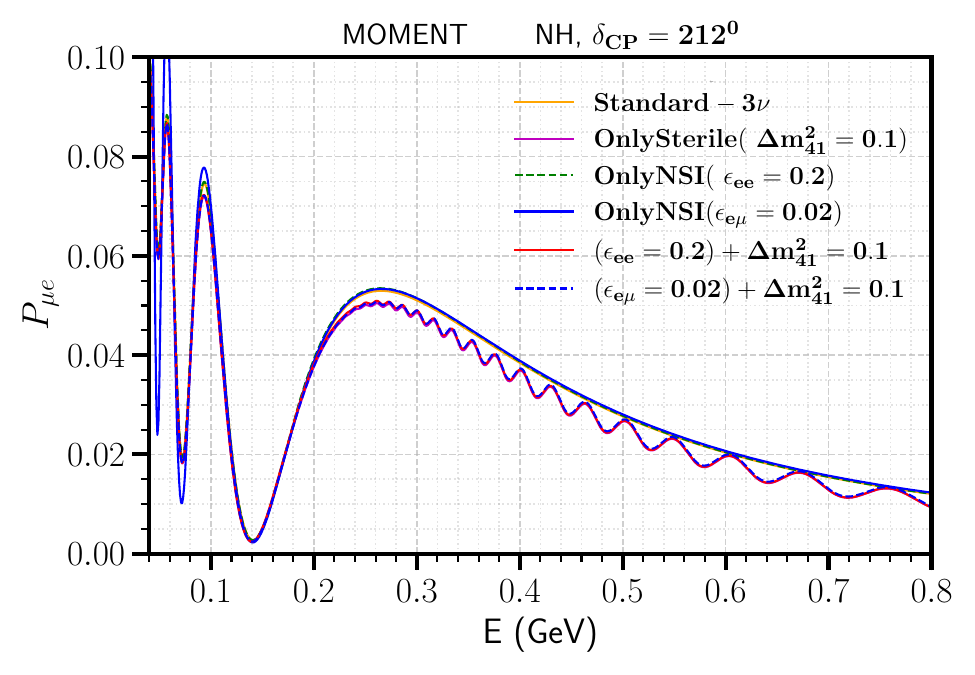}

 \caption{The $\nu_\mu\to\nu_e$ appearance probability 
as a function of neutrino energy for DUNE and MOMENT, for Normal Hierarchy and $\delta_{\rm CP}=212^\circ$. Six scenarios are shown: standard three-flavor, sterile-only with $\Delta m_{41}^2=0.1~\mathrm{eV}^2$, NSI-only with $\varepsilon_{ee}=0.2$ (green dashed) and $\varepsilon_{e\mu}=0.02$ (blue solid), and combined NSI+sterile for each (red solid and blue dashed).}
    \label{prob}
    \end{center}

\end{figure}
Figure~\ref{prob} shows $P_{\mu e}$ as a function of neutrino energy for DUNE and MOMENT at $\delta_{\rm CP}=212^\circ$, normal mass hierarchy, and $\Delta m_{41}^2=0.1~\mathrm{eV}^2$ for the sterile scenarios.
Six distinct parameter combinations are presented to demonstrate both the individual and collective effects of the parameters across the two probability panels.
At DUNE, the six curves are clearly separated. The diagonal NSI parameter $\varepsilon_{ee}=0.2$ increases the effective matter potential, amplifying the MSW-driven appearance peak, while the off-diagonal parameter $\varepsilon_{e\mu}=0.02$,  in contrast, suppresses $P_{\mu e}$ below the SM curve across the entire energy range. This suppression can be from destructive interference between the flavor-changing NSI matter term and the standard oscillation (SO) terms at $\delta_{\rm CP}=212^\circ$. At a different CP phase, the effect can be constructive. The sterile-only curve (magenta) sits slightly above the SO and carries visible fast oscillations. All six curves are visibly distinct at DUNE, meaning the experiment can distinguish between the scenarios but only with sufficient statistics and energy resolution to separate overlapping effects. At MOMENT the picture is notably different. All NSI curves ($\varepsilon_{ee}$ and $\varepsilon_{e\mu}$)  overlap with the standard oscillation probability. This is a direct consequence of MOMENT's short 150~km baseline, making $P_{\mu e}$ insensitive to both $\varepsilon_{ee}$ and $\varepsilon_{e\mu}$ at the probability level. Only the probability curves with sterile neutrino deviate from the SO through the fast-oscillating terms, which are visible even at MOMENT's low energies.

This insensitivity of MOMENT to NSI  makes it a valuable complement to DUNE; it can constrain the sterile mixing and $\delta_{\rm CP}$ more precisely, independent of any NSI assumption, providing an advantage that breaks the NSI-sterile parameter degeneracies.
\subsection{Bounds on Sterile parameters}
\label{sterilebound}
\begin{figure}[h]
\begin{center}
\includegraphics[width=54mm, height=50mm]{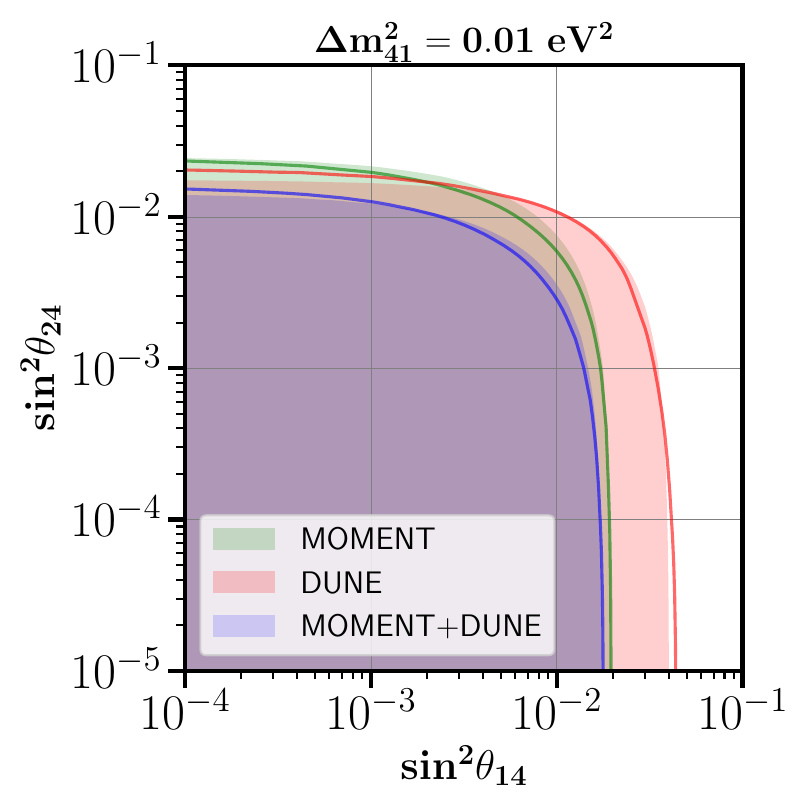}
\includegraphics[width=54mm, height=50mm]{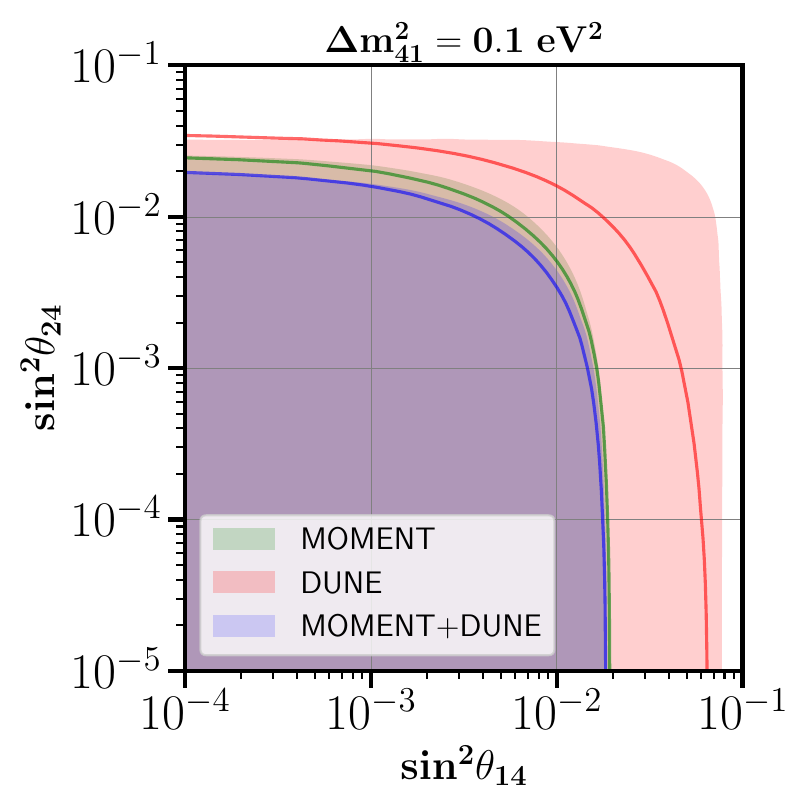}
\includegraphics[width=54mm, height=50mm]{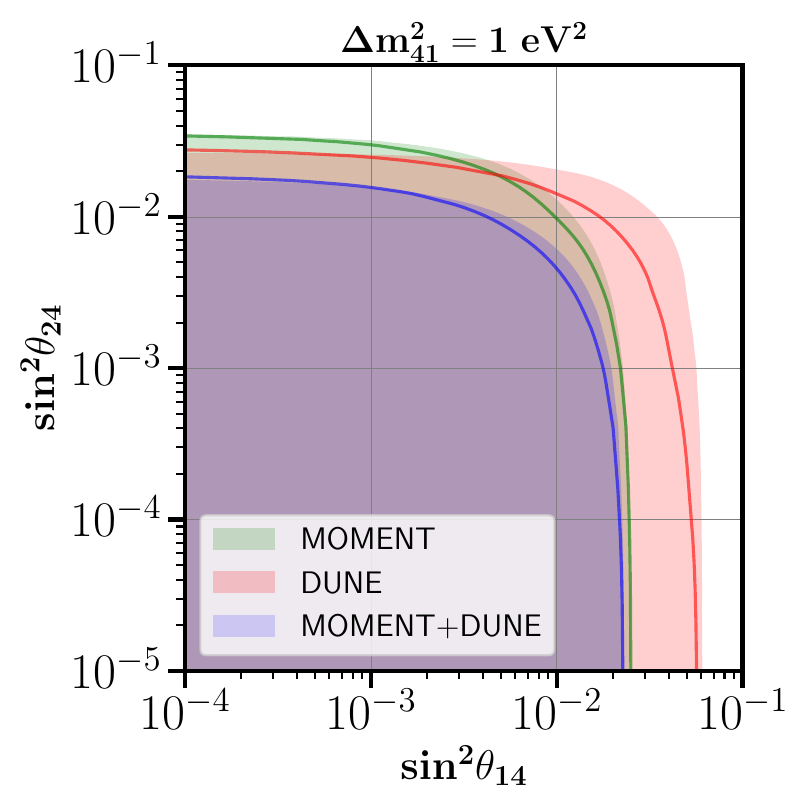}

 \caption{Sensitivity to sterile parameter space in the $\sin^2 \theta_{14}-\sin^2 \theta_{24}$ plane with fixed $\Delta m^2_{41}$ values. All the contours are drawn at $95 \%$ C.L.  The line contours represent the sterile-only scenario, while the filled contours depict the sterile scenario with  NSI. }
    \label{sterile-bounds}
    \end{center}

\end{figure}

Figure~\ref{sterile-bounds} presents the projected $95\%$ C.L. constraints in the $\sin^2\theta_{14}-\sin^2\theta_{24}$ plane for three representative benchmark values of the sterile mass-squared splitting, $\Delta m_{41}^2 = 0.01$, $0.1$, and $1~{\rm eV}^2$. The sensitivities are shown for MOMENT, DUNE, and their synergy, MOMENT+DUNE. For each experimental configuration, the line contours correspond to the pure sterile-neutrino scenario, while the filled contour regions represent the constraints obtained after simultaneously incorporating and marginalizing over the NSI parameters $\varepsilon_{ee}$, $\varepsilon_{e\mu}$. The phase $\phi_{e\mu}$ is marginalized over its full range of $360^\circ$. Across all three panels, the combined MOMENT+DUNE analysis provides the most stringent constraints, with MOMENT individually setting stronger bounds on $\sin^2\theta_{14}$ while DUNE is more 
sensitive to $\sin^2\theta_{24}$. A notable feature visible in all three panels is that the bounds on $\sin^2\theta_{24}$ are consistently two to three times more stringent than those on $\sin^2\theta_{14}$ at any fixed $\Delta m_{41}^2$. This asymmetry arises from the different ways the two angles appear in the oscillation probabilities. $\theta_{14}$ enters only the appearance channel, while  $\theta_{24}$ entering both  appearance and disappearance channels.  

 In the absence of NSI, the contours include smaller regions of the parameter space. This nature is consistent for DUNE+MOMENT, followed by MOMENT and DUNE. The improvements on the bounds obtained from the combined analysis demonstrate the advantage of combining experiments with different oscillation baselines and energy regimes when probing active-sterile neutrino mixing. While DUNE benefits from its long baseline and significant matter effects, MOMENT provides an independent measurement with a different energy spectrum and cleaner, low-energy, highly intense neutrino fluxes. 

\subsection{Bounds on NSI parameters}
\label{nsibound}
\begin{figure}[h]
\begin{center}
\includegraphics[width=60mm, height=54mm]{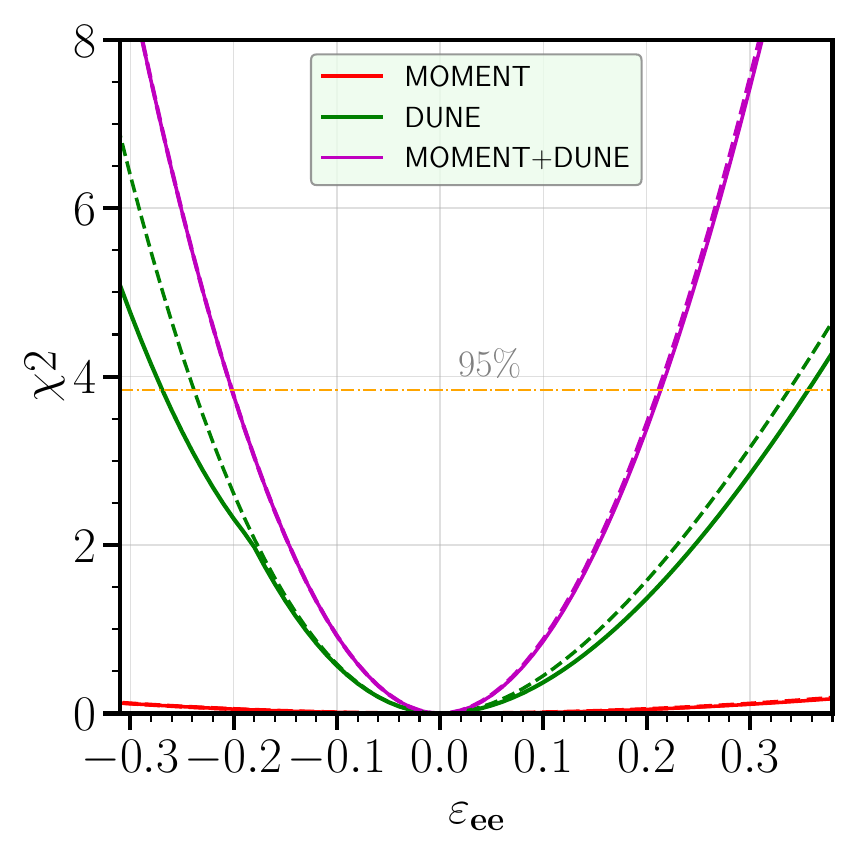}
\quad 
\includegraphics[width=60mm, height=54mm]{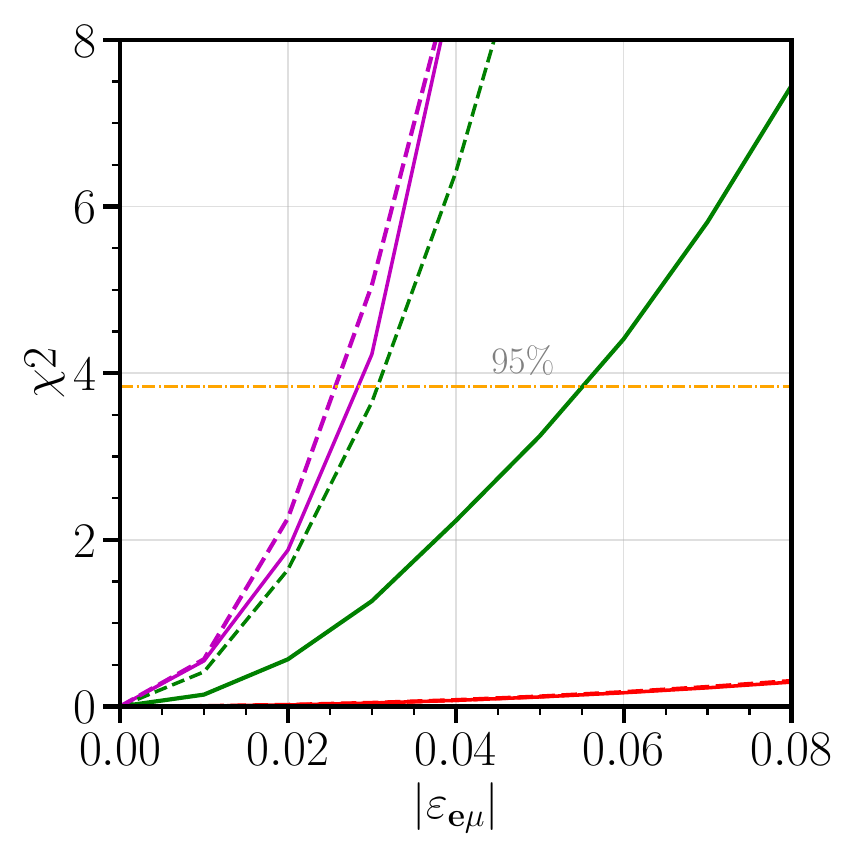}\\

 \caption{Sensitivity to NSI parameters  $\varepsilon_{ee}$ and $\varepsilon_{e\mu}$. The horizontal line is drawn at $95 \%$ C.L. The left (right) panels illustrate bounds on individual experiments. The dashed lines curve to represent the only NSI scenario, while the solid curves present  NSI in the presence of an additional sterile neutrino, where we have considered $\Delta m_{41}^2 =1~\mathrm{eV}^2$ in both the panels.}
    \label{nsi-bounds}
    \end{center}

\end{figure}

Figure~\ref{nsi-bounds} shows the projected sensitivities to the NSI parameters for MOMENT, DUNE, and their synergetic set up. The left and right panels display the $\Delta\chi^2$ profiles as functions of $\varepsilon_{ee}$ and $|\varepsilon_{e\mu}|$, respectively.   $\phi_{e\mu}$ is marginalized over its full range. The  dashed and solid curves correspond to the NSI-only and NSI+sterile scenarios, with the sterile mass squared difference  $\Delta m_{41}^2$ fixed at $1~\mathrm{eV}^2$. 
The inclusion of sterile neutrino mixing results in a modest reduction of sensitivity. DUNE alone constrains $\varepsilon_{ee} \in [-0.24,\,+0.33]$ in the NSI-only scenario, consistent with the expected sensitivity of long-baseline experiments. The combination of MOMENT+DUNE tightens the bounds to approximately $\varepsilon_{ee} \in [-0.20,\,+0.21]$ at 95\% CL. Although MOMENT's sensitivity to $\varepsilon_{ee}$  is limited, a direct consequence of its short 150~km  baseline suppresses any matter-induced NSI dependence. Consequently, it can constrain the sterile-neutrino parameters without NSI contamination, thereby breaking the NSI-sterile degeneracy.

A different picture emerges for the off-diagonal parameter $|\varepsilon_{e\mu}|$, shown in the right panel. In this case, the sensitivity curves exhibit a significantly stronger dependence on the presence of sterile neutrino mixing.  This effect is particularly pronounced for DUNE analysis. While the NSI-only scenario yields stringent constraints on $|\varepsilon_{e\mu}|$, the inclusion of sterile parameters systematically reduces the sensitivity, leading to a visible shift in the $\chi^2$ profile. The NSI flavor-changing term has its own phase $\phi_{e\mu}$ along with $\delta_{24}$ and the standard CP phase. Thus, there exists a large manifold of sterile and NSI combinations that produce nearly identical probabilities, substantially enlarging the degenerate regions, thereby weakening the exclusion power.


\subsection{CP-Violation Sensitivity in the Presence of Sterile Neutrinos and NSI}

\begin{figure}[h]
\begin{center}
\includegraphics[width=54mm, height=51mm]{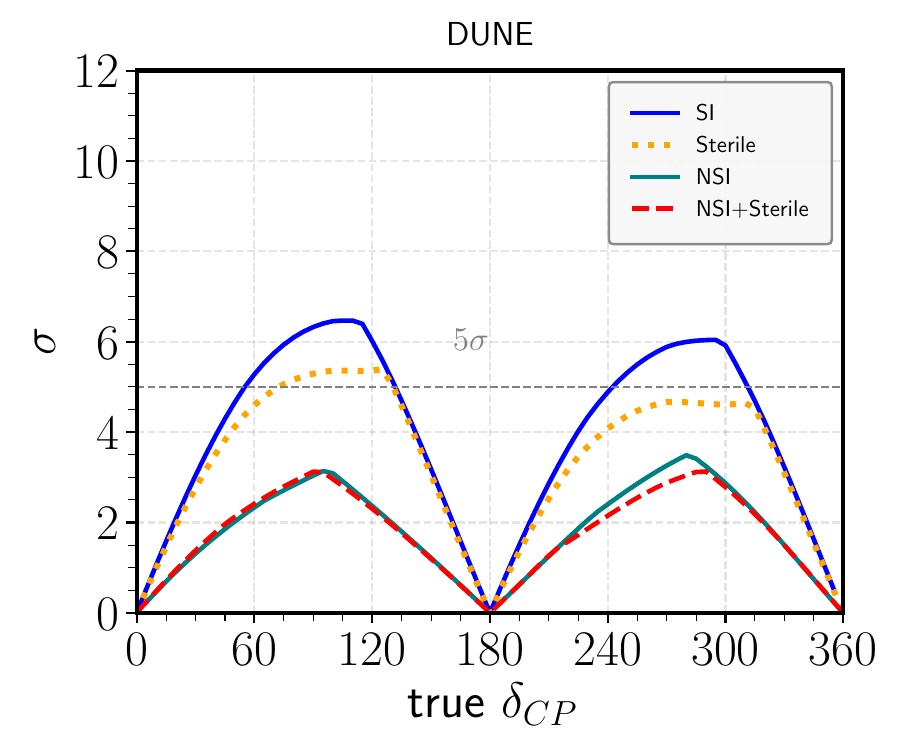}
\includegraphics[width=54mm, height=51mm]{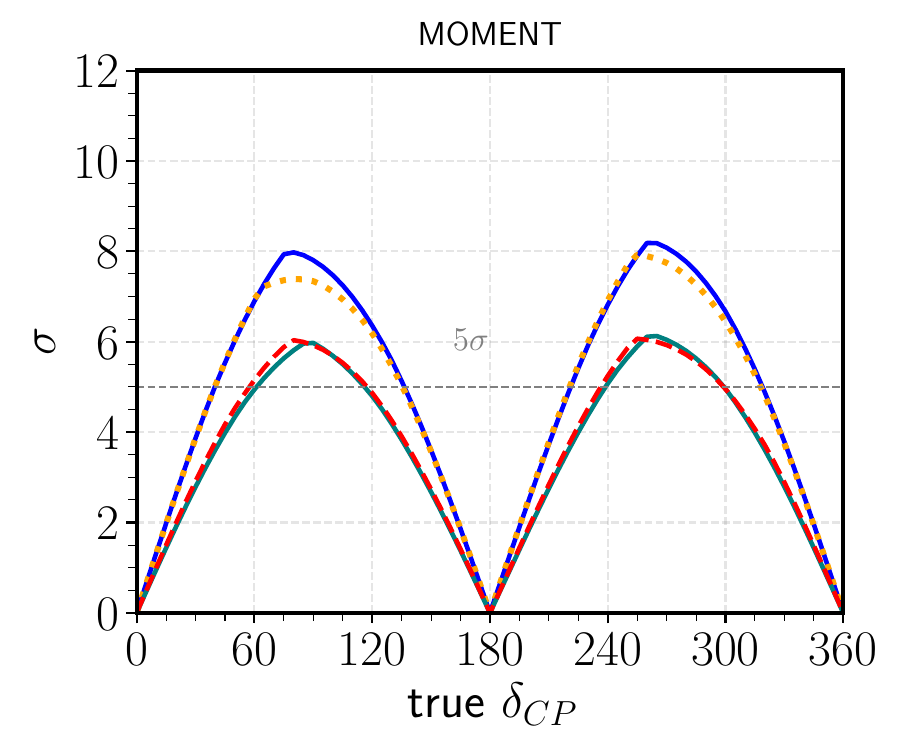}
\includegraphics[width=54mm, height=51mm]{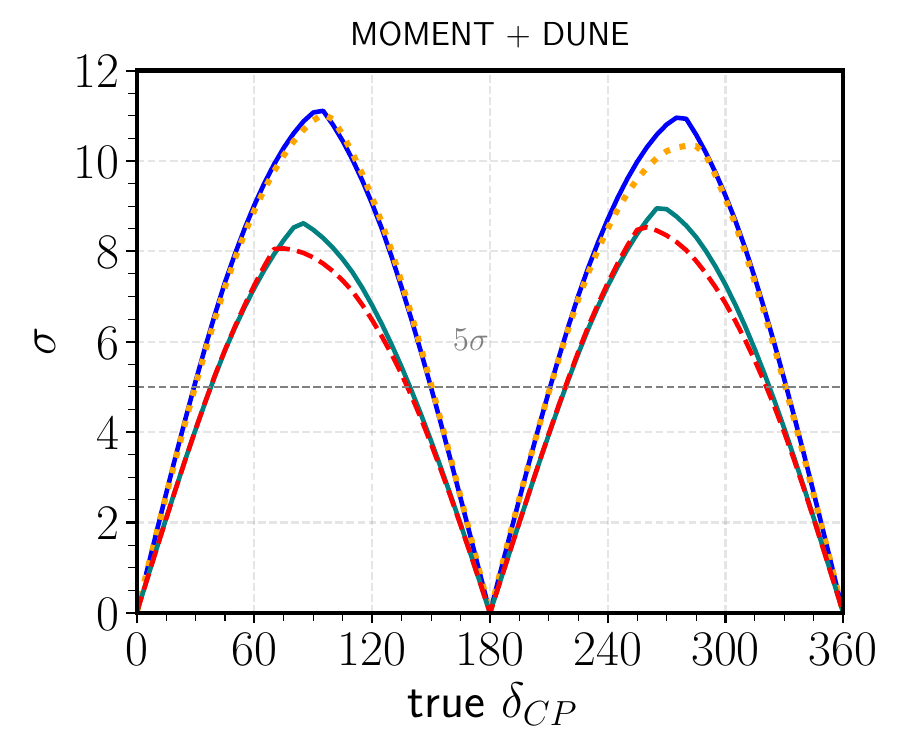}
 \caption{CPV sensitivity in the standard 3, $3+1$ sterile frameworks, in the presence of NSI parameters  $\varepsilon_{ee}$ and $\varepsilon_{e\mu}$, and a combined presence of both $3+1$ sterile with NSI parameters. $\Delta m_{41}^{2}=1~{\rm eV}^{2}$ for all the panels.}
    \label{cpv-sen}
    \end{center}

\end{figure}

One of the primary goals of current and future long-baseline neutrino oscillation experiments is to   measure the leptonic CP violation. In the standard three-flavor framework, CP violation arises from the complex phase $\delta_{\rm CP}$ present in the PMNS mixing matrix. A non-zero CP-violating phase leads to different oscillation probabilities for neutrinos and antineutrinos,
\begin{equation}
P(\nu_\alpha \rightarrow \nu_\beta)
\neq
P(\bar{\nu}_\alpha \rightarrow \bar{\nu}_\beta),
\end{equation}
providing a direct signature of CP violation in the lepton sector. 
The situation becomes more complex in the presence of physics beyond the Standard Model. In the $3+1$ sterile-neutrino framework, additional CP-violating phases are introduced through the extended mixing. Similarly, NSI introduces new complex phases associated with the off-diagonal parameters. These additional sources of CP violation can interfere with the standard phase $\delta_{\rm CP}$, potentially altering the CP sensitivity of oscillation experiments and generating parameter degeneracies. It is therefore essential to investigate how sterile neutrinos, NSI, and their combination affect the capability of future experiments to establish CP violation.  

Figure~\ref{cpv-sen} illustrates the impact of sterile neutrinos and non-standard interactions on the CP-violation sensitivity of DUNE, MOMENT, and their combination. We show four scenarios: standard three-flavor (SI, blue solid), 3+1 sterile only (orange dotted), NSI only (green solid), and NSI+sterile (red dashed), all for $\Delta m_{41}^2 = 1~\mathrm{eV}^2$.

A comparison among the different scenarios reveals that the standard three-flavor framework consistently yields the highest CP sensitivity. The introduction of sterile neutrino mixing produces a moderate reduction in significance, while the inclusion of NSI leads to a substantially larger degradation. This behavior is particularly evident for DUNE, where strong matter effects enhance the impact of propagation NSI. The near overlap of the NSI-only and NSI+sterile curves further indicates that the loss of sensitivity is driven primarily by NSI rather than by sterile mixing. The mechanism is the $\phi_{e\mu}-\delta_{\rm CP}$ degeneracy: marginalizing over the free NSI phase $\phi_{e\mu}$ allows a suitable $(|\varepsilon_{e\mu}|,\,\phi_{e\mu})$ combination to mimic the CP-violating effects. MOMENT is more robust to this degeneracy because the NSI matter effect is minimal due to its small baseline.

Among the three experimental configurations, the combined DUNE+MOMENT analysis exhibits the best performance throughout the entire $\delta_{\rm CP}$ range. The combination significantly retains a sensitivity exceeding the $5\sigma$ level over a broad range of true $\delta_{\rm CP}$ values even in the presence of both sterile neutrinos and NSI. Overall, the results indicate that while both sterile neutrinos and NSI weaken CP-violation sensitivity, the dominant limitation arises from NSI effects.

 We have also probed the robustness of the DUNE+MOMENT complementarity against the assumed experimental systematic uncertainties; the corresponding results are presented in Appendix \ref{app:statistics}.
\subsection{Correlation among the standard, sterile, and NSI parameters}

\begin{figure}[h]
\begin{center}
\includegraphics[width=160mm, height=160mm]{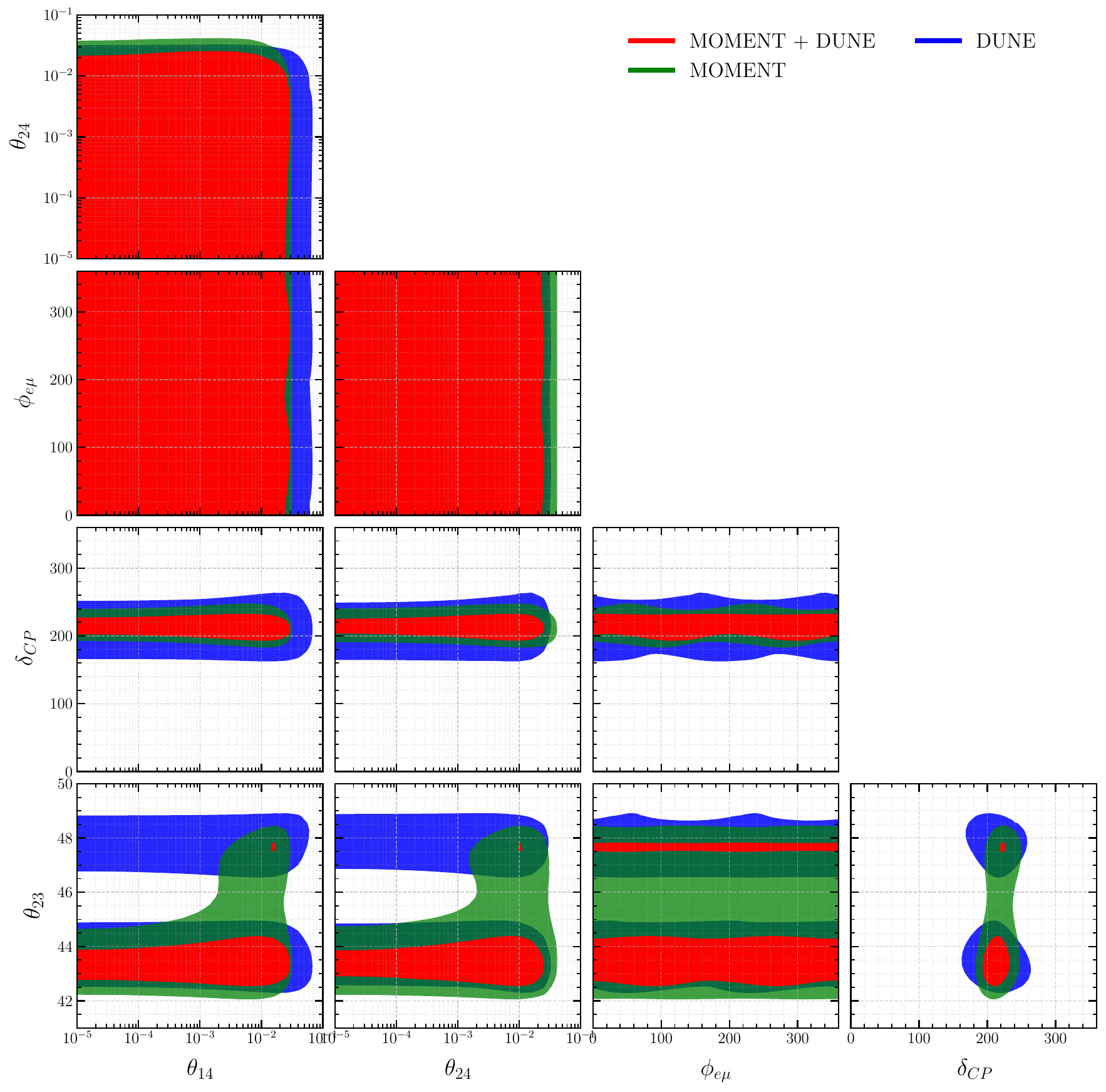}
 \caption{Two-dimensional allowed regions for different combinations of oscillation parameters (both standard and new physics) at $95\%$~C.L.~ for DUNE (blue), MOMENT (green), and MOMENT+DUNE (red), shown for $\Delta m_{41}^2 = 1~\mathrm{eV}^2$. The displayed parameters are $\sin^2\theta_{14}$, $\sin^2\theta_{24}$, the NSI phase $\phi_{e\mu}$, the standard CP phase $\delta_{\rm CP}$, and the atmospheric angle $\theta_{23}$. }
    \label{correl}
    \end{center}

\end{figure}

To further explore the interplay among the sterile neutrino mixing and non-standard interaction parameters, we present in Fig.~\ref{correl} the two-dimensional allowed regions among the parameters $\theta_{14}$, $\theta_{24}$, $\phi_{e\mu}$, $\delta_{\rm CP}$, and $\theta_{23}$. The analysis is performed for $\Delta m_{41}^{2}=1~{\rm eV}^{2}$, and $\Delta m_{31}^{2}$ is marginalized over its allowed $3\sigma$ range. The remaining oscillation parameters are treated as free parameters with appropriate priors. All contours correspond to the $95\%$ confidence level.

An important outcome of the figure is the absence of strong correlations involving the NSI phase $\phi_{e\mu}$. This can be observed from the nearly rectangular regions in the $(\theta_{14},\phi_{e\mu})$ and $(\theta_{24},\phi_{e\mu})$ planes, where the allowed ranges of the sterile mixing angles remain largely unchanged across the full interval of $\phi_{e\mu}$. Such behaviors indicate that the current experimental configurations possess limited sensitivity to the NSI phase and that the sterile mixing parameters are effectively constrained independent of $\phi_{e\mu}$. The broad allowed range of $\phi_{e\mu}$ further suggests that the dominant sensitivity arises primarily from the magnitude of the NSI coupling rather than its associated complex phase.

A similar pattern is observed in the planes involving the standard CP-violating phase $\delta_{\rm CP}$. The contours in the $(\theta_{14},\delta_{\rm CP})$ and $(\theta_{24},\delta_{\rm CP})$ parameter spaces exhibit only weak distortions and remain approximately aligned with the coordinate axes near the true value $\delta_{\rm CP}=212^\circ$. This indicates that the determination of the sterile mixing angles is not strongly affected by variations of $\delta_{\rm CP}$. Conversely, the allowed range of $\delta_{\rm CP}$ remains relatively confined around its true value even after marginalizing over the sterile and NSI sectors. 

The most pronounced correlations are observed in the planes involving the atmospheric mixing angle $\theta_{23}$. In both the $(\theta_{14},\theta_{23})$ and $(\theta_{24},\theta_{23})$ parameter spaces, the allowed regions split into two disconnected branches corresponding to the lower- and higher-octant solutions of $\theta_{23}$. This behavior originates from the well-known octant degeneracy of the atmospheric sector. Since the $\nu_\mu\rightarrow\nu_e$ appearance probability depends on both $\theta_{23}$ and the sterile mixing parameters, changes in the atmospheric mixing angle can be partially compensated by corresponding shifts in $\theta_{14}$ and $\theta_{24}$. Consequently, distinct combinations of $(\theta_{23},\theta_{14},\theta_{24})$ may produce nearly identical oscillation spectra, leading to multiple local minima in the fit. The persistence of the two-branch structure demonstrates that neither sterile-neutrino mixing nor NSI effects completely resolve the octant ambiguity. Instead, the atmospheric sector remains the dominant source of degeneracy in the combined sterile+NSI parameter space.

Another notable result is the different behavior exhibited by the sterile mixing angles $\theta_{14}$ and $\theta_{24}$. The allowed regions generally show a stronger constraint in the $\theta_{24}$ direction than in $\theta_{14}$.  This behavior can be understood from the structure of the oscillation probabilities in the $3+1$ framework. While both angles contribute to the appearance channel through the combination $s_{14}s_{24}$, the mixing angle $\theta_{24}$ additionally enters directly into the $\nu_\mu$ disappearance probability. As a result, long-baseline experiments constrain $\theta_{24}$ through both appearance and disappearance measurements, whereas $\theta_{14}$ is probed primarily through appearance data. The larger statistics and precision of the muon-disappearance channel therefore lead to a comparatively stronger sensitivity to $\theta_{24}$, resulting in tighter allowed regions along this direction. A more specific correlations involving $|\epsilon_{e\mu}|,\delta_{24}$, and $\delta_{\rm CP}$ are examined separately in Sec.\ref{phase_and_NSI}.

Comparing the individual experiments with their combination, it is clear that the combined configuration consistently provides the smaller allowed regions throughout the parameter space. The improvement is not merely statistical but arises from the complementary sensitivities of both experiments. The combination, therefore, helps disentangle the contributions of sterile mixing, CP violation, and NSI effects, leading to a substantial reduction of the allowed parameter space.

\subsection{Phases and NSI Correlations}
\label{phase_and_NSI}
\begin{figure}[h]
\begin{center}
\includegraphics[width=54mm, height=50mm]{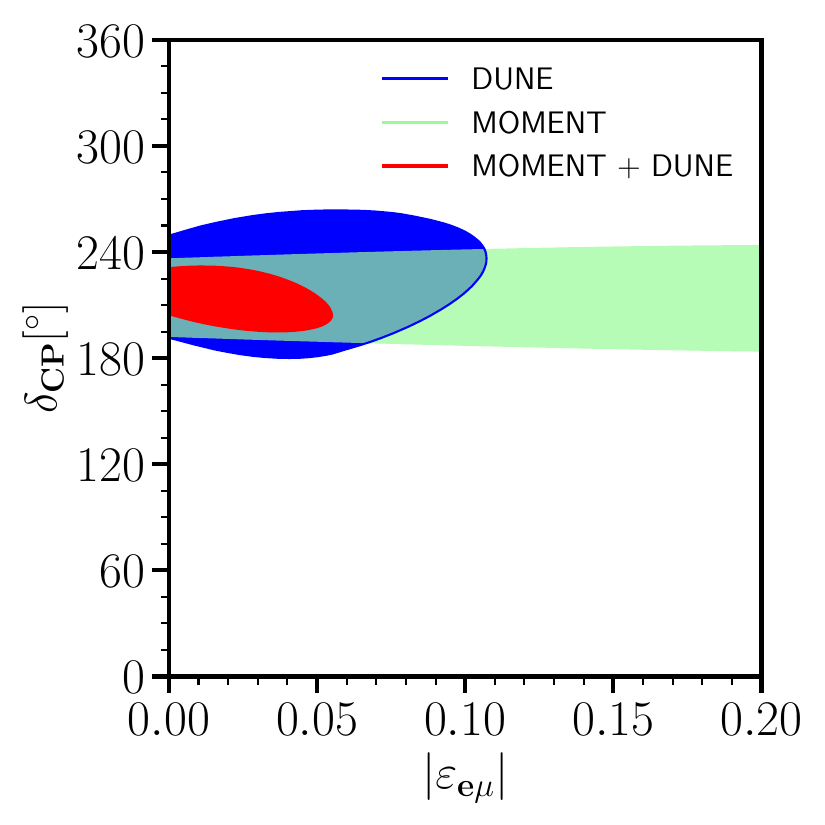}
\includegraphics[width=54mm, height=50mm]{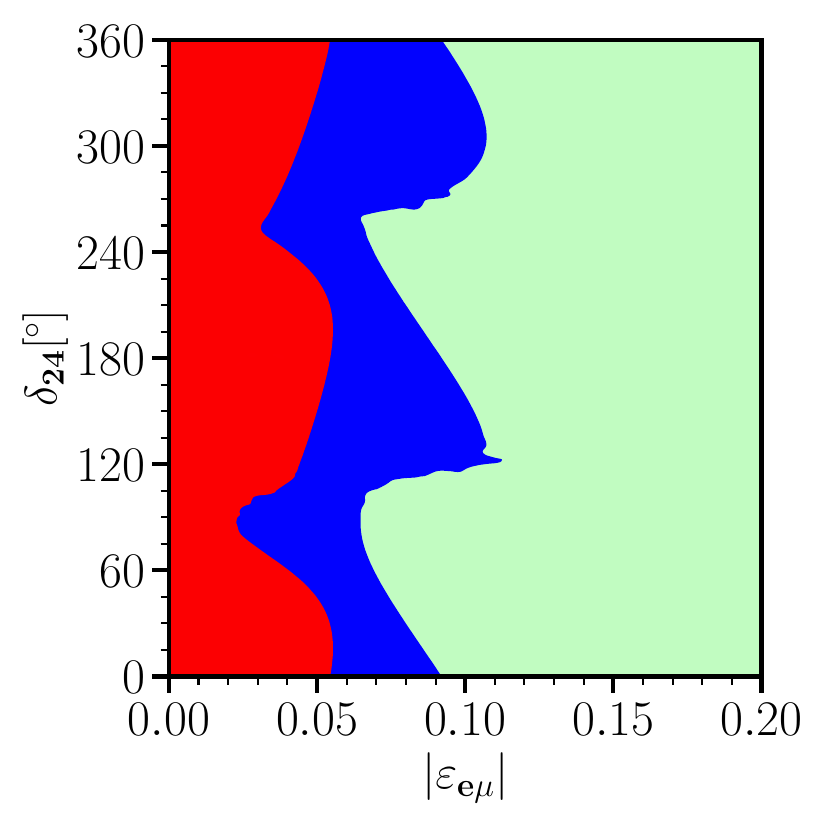}
\includegraphics[width=54mm, height=50mm]{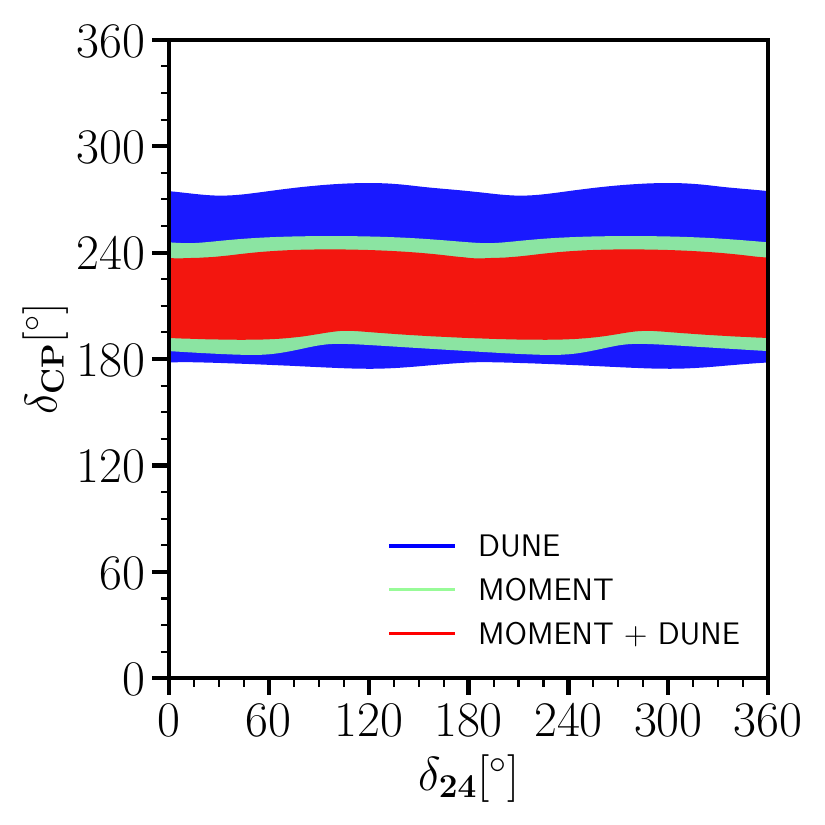}
 \caption{Two-dimensional allowed regions at $95\%$ C.L. in the $(|\epsilon_{e\mu}|,\delta_{\rm CP})$, $(|\epsilon_{e\mu}|,\delta_{24})$, and $(\delta_{24},\delta_{\rm CP})$ planes for DUNE, MOMENT, and MOMENT+DUNE, for $\Delta m_{41}^{2}=1~{\rm eV}^{2}$. All remaining parameters are profiled over according to the marginalization procedure described in Sec.\ref{Experimental_Setup}. }
    \label{newcorr}
    \end{center}

\end{figure}

The correlation analysis in Fig. \ref{correl} provides a broad view of the interplay among the sterile, NSI, and standard oscillation parameters. To further clarify the origin of the $\delta_{\rm CP}$ narrowing in the combined DUNE+MOMENT analysis, we examine the correlations among the flavor-changing NSI magnitude $|\epsilon_{e\mu}|$, the sterile phase $\delta_{24}$, and the standard CP phase $\delta_{\rm CP}$. We consider $\Delta m_{41}^{2}=1~{\rm eV}^{2}$, while profiling over the remaining free parameters with the same prescription used in the main analysis.

Figure~\ref{newcorr} shows the allowed regions in the $(|\epsilon_{e\mu}|,\delta_{\rm CP})$ , $(|\epsilon_{e\mu}|,\delta_{24})$, and $(\delta_{24},\delta_{\rm CP})$ planes for DUNE, MOMENT, and their combination. The $(|\epsilon_{e\mu}|,\delta_{\rm CP})$ projection exhibits a clear correlation between the allowed NSI magnitude and the determination of the standard CP phase. In particular, the combined DUNE+MOMENT configuration substantially reduces the allowed range of $|\epsilon_{e\mu}|$, accompanied by a narrower allowed region in $\delta_{\rm CP}$. This demonstrates that the improved determination of the standard CP phase is closely connected to the stronger constraint on the flavor-changing NSI sector.

In contrast, the $(|\epsilon_{e\mu}|,\delta_{24})$ projection does not exhibit a pronounced phase-correlated structure. The $(\delta_{24},\delta_{\rm CP})$ plane shows a very weak dependence of the allowed $\delta_{\rm CP}$ region on the sterile phase. The allowed band remains relatively stable as $\delta_{24}$ is varied over its full range.  These results indicate that, for the benchmark parameter space considered here, the sterile phase $\delta_{24}$ does not generate a strong additional degeneracy with either $\delta_{\rm CP}$ or $|\epsilon_{e\mu}|$.

\section{Conclusion}
\label{coclusion}
In this work, we have investigated for the first time the combined effects of non-standard interaction and sterile neutrinos in the context of two upcoming experiments: the long-baseline experiment DUNE and the medium-baseline MOMENT experiment. We found that the simultaneous presence of these two effects significantly modifies the sensitivities of these experiments, and our findings are summarized below.

The results presented in Figs.~\ref{sterile-bounds} and ~\ref{nsi-bounds} demonstrate a clear interplay between sterile neutrino mixing and non-standard interactions. The presence of NSI weakens the exclusion limits on the sterile mixing angles, and analogously, the sterile-neutrino mixing reduces the sensitivity to NSI parameters. The impact is most significant for the off-diagonal parameter $\varepsilon_{e\mu}$, which exhibits a strong correlation with the active-sterile mixing sector. In contrast, the diagonal parameter $\varepsilon_{ee}$ remains moderately affected. Overall, the combined DUNE+MOMENT analysis provides the most stringent and stable constraints in all scenarios considered, illustrating the importance of multi-experiment analyses for resolving sterile-NSI degeneracies and obtaining robust bounds on the parameters of these BSM scenarios.

The inclusion of sterile neutrinos and NSI reduces the CP-violation discovery reach, with NSI producing a more pronounced effect than sterile mixing alone. While the individual sensitivities of DUNE and MOMENT are affected by these new-physics scenarios, their combination substantially improves overall performance and reduces associated parameter degeneracies, thereby providing the most robust determination of leptonic CP violation, even in the simultaneous presence of sterile neutrinos and non-standard interactions.

Overall, the correlation plots indicate that including sterile neutrino mixing and NSI does not introduce strong additional phase degeneracies in the parameter space considered. In particular, the NSI phase $\phi_{e\mu}$ exhibits only weak correlations with both the sterile mixing angles and the standard CP phase $\delta_{\rm CP}$, suggesting that these parameters can be constrained largely independently without much effect from them. The most prominent ambiguity instead arises from the atmospheric mixing angle $\theta_{23}$, whose lower- and higher-octant solutions remain allowed even after combining DUNE and MOMENT data, though the allowed region is very small in the upper octant. Consequently, the dominant limitation on the determination of sterile and NSI parameters originates from the atmospheric sector rather than from the additional CP phases introduced by new physics. These results highlight the importance of improving the precision on $\theta_{23}$ and resolving its octant ambiguity, as such improvements are likely to have a greater impact on future sterile+NSI studies than further constraints on the NSI phase itself.

The correlation analysis further shows that the dominant correlations relevant for the CP determination arise from the flavor-changing NSI sector. The $(|\varepsilon_{e\mu}|,\delta_{\rm CP})$ projection demonstrates that the stronger constraint on $|\varepsilon_{e\mu}|$ in the combined DUNE+MOMENT analysis is accompanied by a narrower allowed $\delta_{\rm CP}$ region. In contrast, the sterile phase $\delta_{24}$ exhibits only weak correlations with both $\delta_{\rm CP}$ and $|\varepsilon_{e\mu}|$ for the benchmark considered. The remaining prominent ambiguity is associated with the atmospheric angle $\theta_{23}$ and its octant.

Taken together, these results establish three conclusions relevant to the experimental program of the coming decade. First, NSI is a more adverse source of parameter degeneracy than sterile-neutrino mixing at eV-scale $\Delta m_{41}^2$, because the sterile signal is energy-averaged at long baselines while the NSI phase remains degenerate with $\delta_{\rm CP}$. Second, no single experiment can robustly measure $\delta_{\rm CP}$ or place definitive bounds on either BSM sector when the other is simultaneously unknown; the combination of a matter-rich and a near-vacuum experiment is the minimal strategy needed to disentangle them.  Third, reducing the uncertainty in NSI parameters through external constraints, such as CEvNS measurements or dedicated short-baseline searches, would have a larger positive impact on CPV discovery reach than any purely statistical improvement in a single long-baseline detector. Future analyses combining various experiments in a joint framework would be a natural extension of the present work.

\vspace*{0.2 true cm}
{\bf Acknowledgments}

\vspace*{0.2 true cm}
SKP thanks the University Grants Commission for the NFOBC fellowship. TG acknowledges the UGC-JRF fellowship for supporting her doctoral work.
\appendix
\section{Dependence on Systematic Uncertainties analysis}
\label{app:statistics}

\begin{figure}[h]
\begin{center}
\includegraphics[width=54mm, height=51mm]{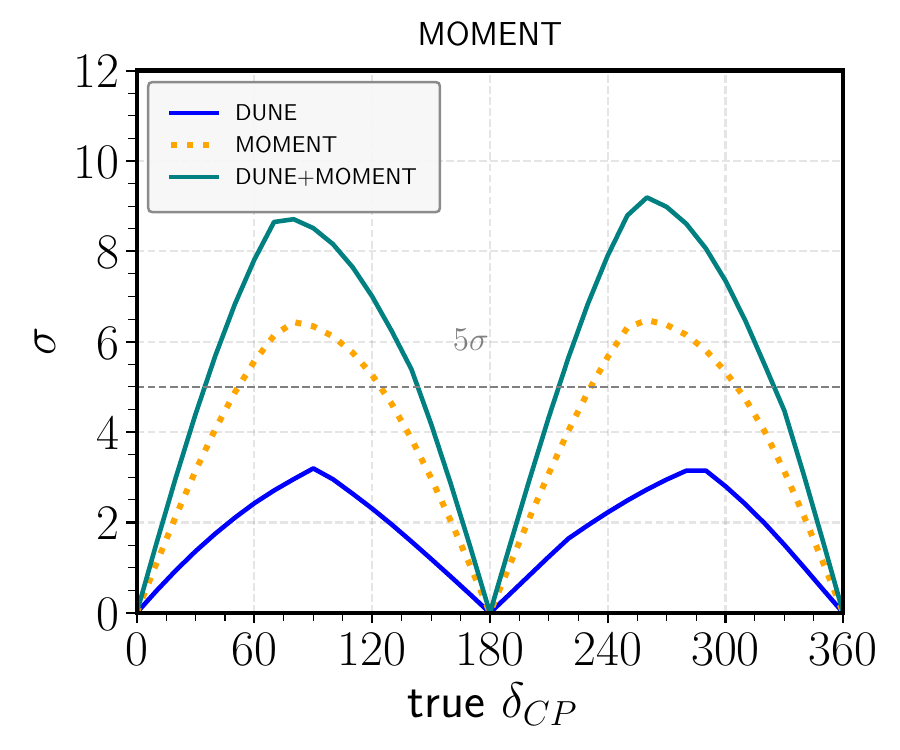}
\includegraphics[width=54mm, height=51mm]{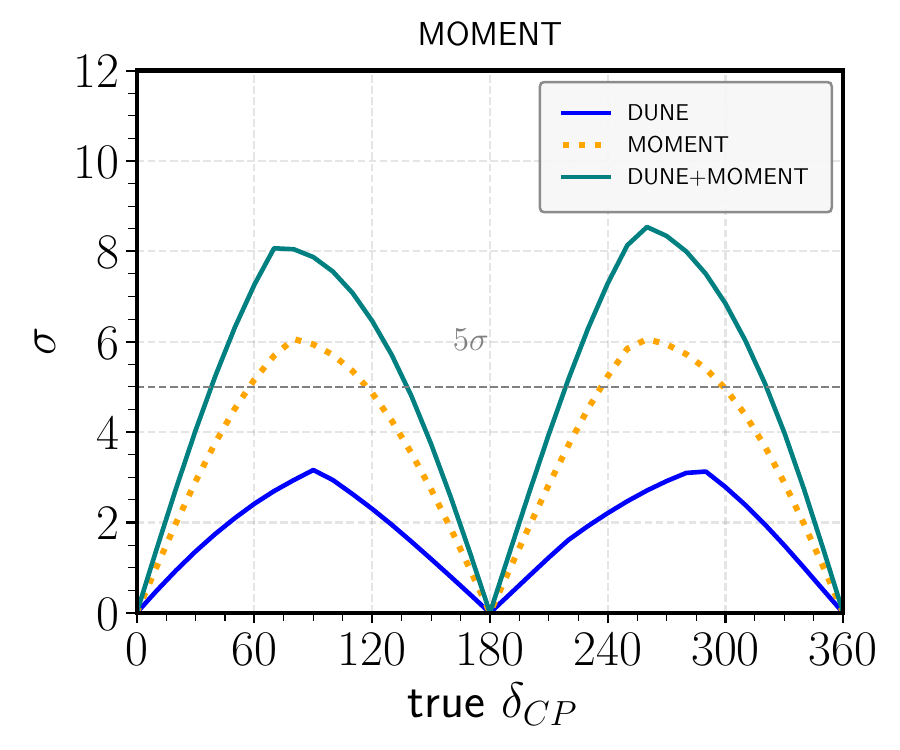}
\includegraphics[width=54mm, height=51mm]{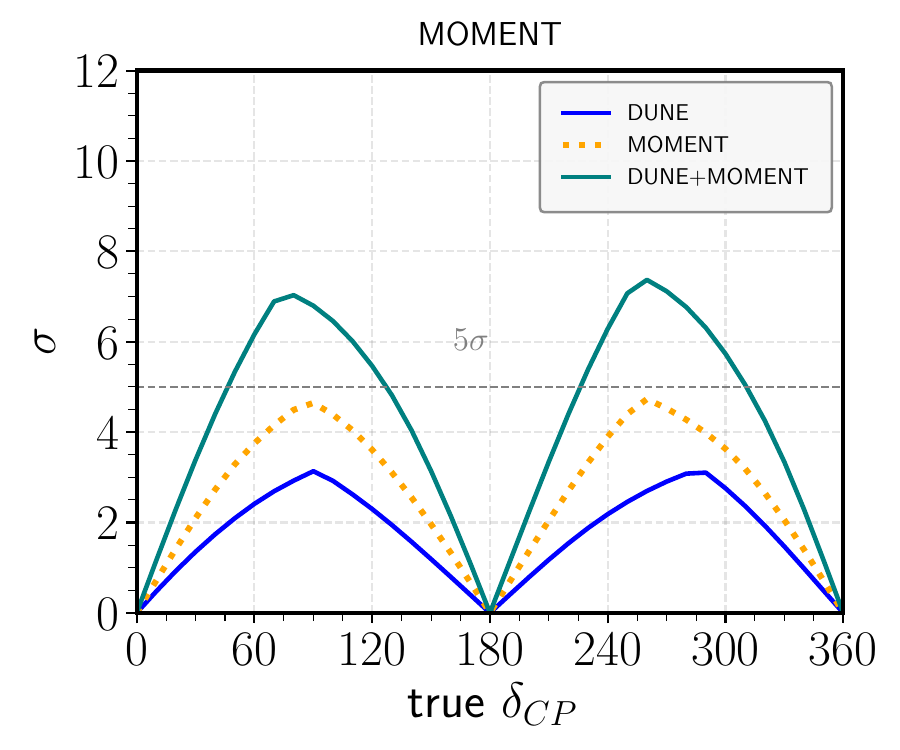}
 \caption{CP-violation sensitivity for DUNE, MOMENT, and their combination under three systematic configurations. From left to right: experimental systematic uncertainties switched off; a common $5\%$ background normalization uncertainty and a common $20\%$ background normalization uncertainty. The comparison shows that the DUNE+MOMENT combination retains enhanced CP sensitivity under all three assumptions.}
    \label{cpv-bgcheck}
    \end{center}

\end{figure}

\begin{figure}[h]
\begin{center}
\includegraphics[width=54mm, height=51mm]{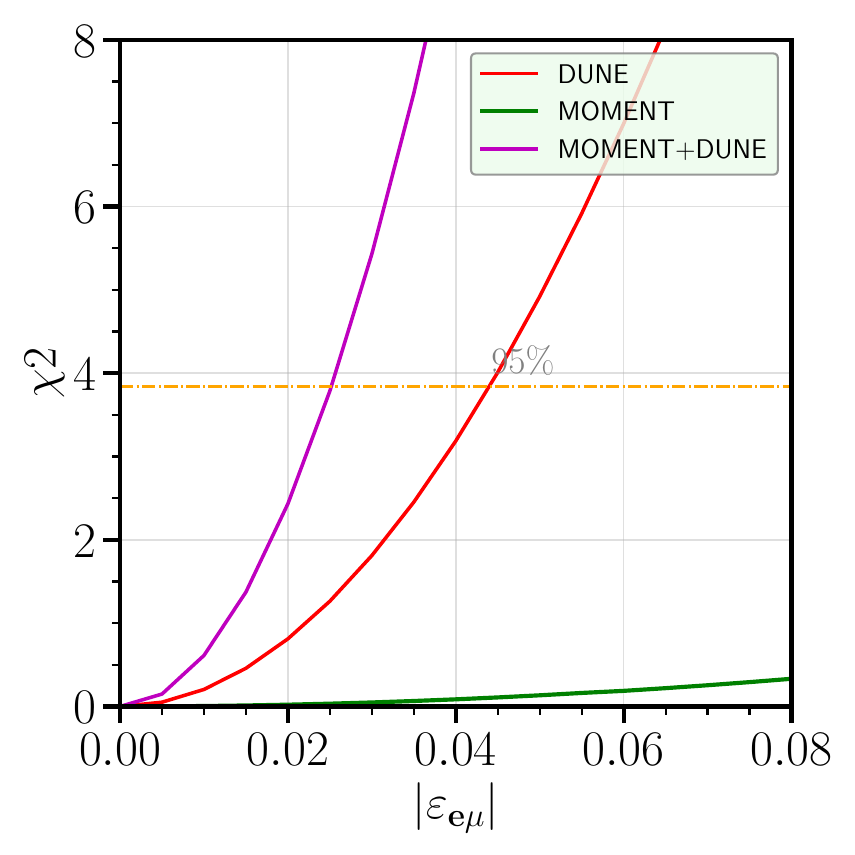}
\includegraphics[width=54mm, height=51mm]{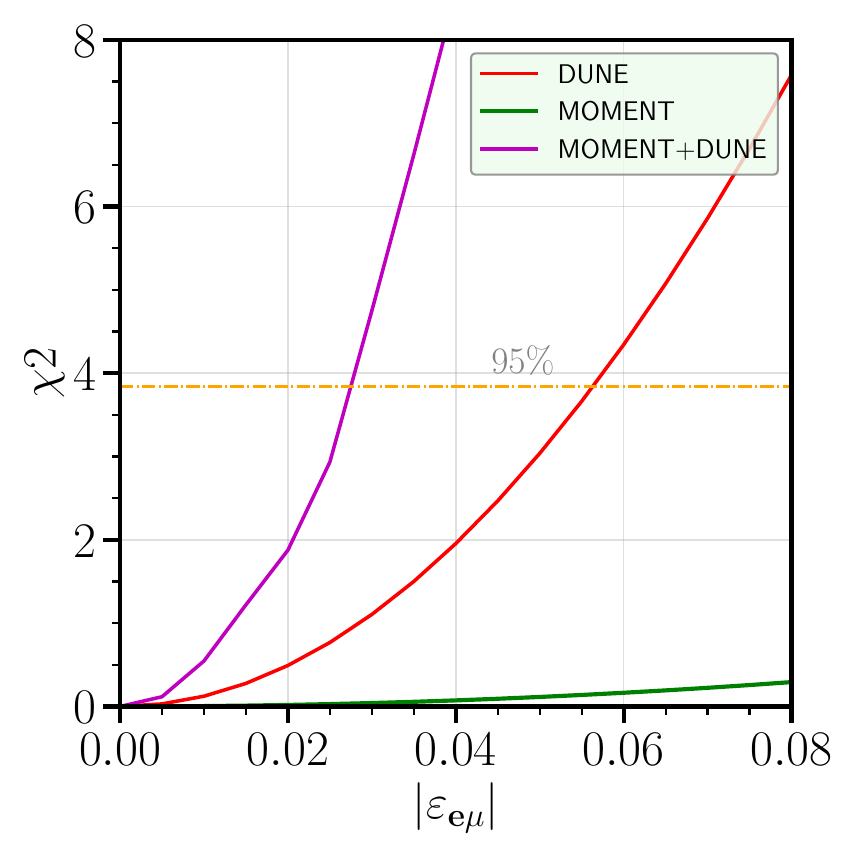}
\includegraphics[width=54mm, height=51mm]{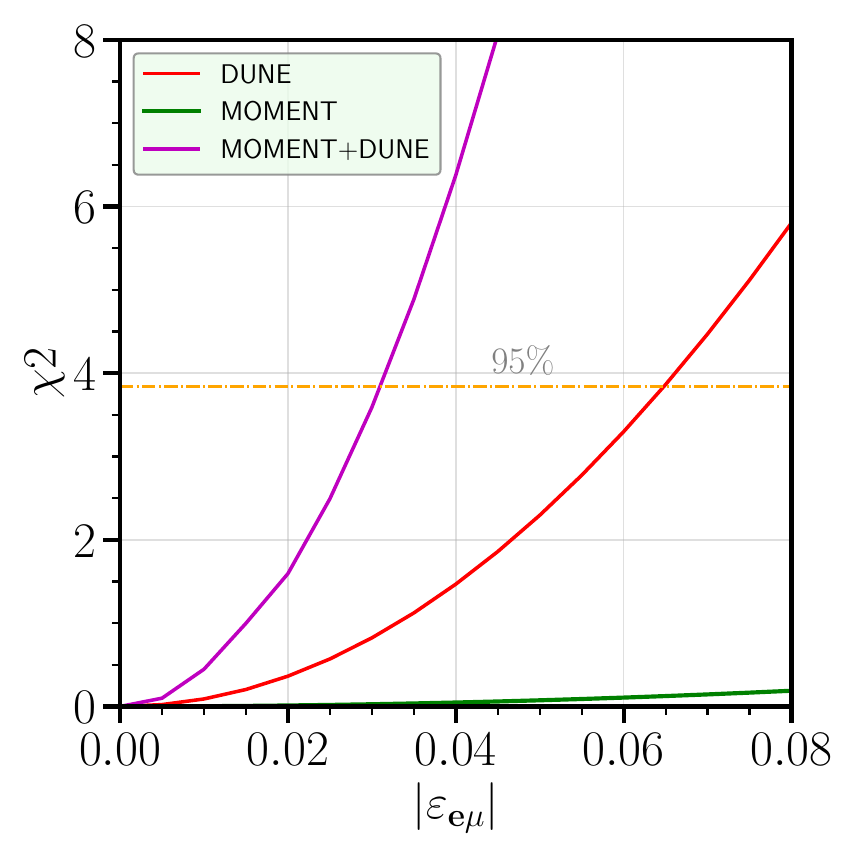}
 \caption{$\chi^2$ sensitivity to $|\varepsilon_{e\mu}|$ at $95\%$ C.L.  From left to right: experimental systematic uncertainties switched off; a common $5\%$ background normalization uncertainty and a common $20\%$ background normalization uncertainty.}
    \label{emu-bgcheck}
    \end{center}

\end{figure}

To test whether the DUNE–MOMENT complementarity is driven by the different assumptions for background uncertainties, we repeated the analysis under various systematic configurations. First, the background normalization uncertainty was set to a common value of $5\%$ and then to $20\%$ for both DUNE and MOMENT. Second, all experimental systematic uncertainties were switched off, corresponding to the statistics-dominated limit. The resulting sensitivities for $|\varepsilon_{e\mu}|$ and CP violation are compared in Figs.~\ref{cpv-bgcheck} and \ref{emu-bgcheck}. While the absolute sensitivities vary with the assumed systematic uncertainties, the combined DUNE+MOMENT configuration consistently retains an improvement over the individual experiments. This confirms that the observed complementarity is not solely a consequence of the smaller background uncertainty assumed for MOMENT.

\bibliography{mybibfile}
\bibliographystyle{ieeetr} 
\end{document}